\documentclass[preprint]{elsarticle}

\usepackage[american]{babel}
\usepackage{graphicx}

\begin{document}

\title{Crystalline misfit-angle implications for solid sliding}

\author[ma]{Nicola Manini\corref{cor1}}
\address[ma]{Dipartimento di Fisica, Universit\`a degli Studi di Milano,\\
Via Celoria 16, 20133 Milano, Italy
}
\cortext[cor1]{Phone: +39 02 50317355; Fax: +39 02 50317482;
  email: nicola.manini@mi.infm.it}

\author[br]{O.M. Braun\corref{cor2}}
\address[br]{Institute of Physics, National Academy of Sciences of Ukraine,\\
46 Science Avenue, Kiev 03028, Ukraine}
\cortext[cor2]{email: obraun.gm@gmail.com}
\date{\today}

\begin{abstract}
For the contact of two finite portions of interacting rigid crystalline
surfaces, we compute the the pinning energy barrier dependency on the
misfit angle and contact area.
This simple model allows us to investigate a broad contact-size and angular
range, thus obtaining the statistical properties of the energy barriers
opposing sliding for a single asperity.
These data are used to generate the distribution of static frictional
thresholds for the contact of polycrystals, as in dry or even lubricated
friction.
This distribution is used as the input of a master equation to predict the
sliding properties of macroscopic contacts.
\end{abstract}

\begin{keyword}
tribology
\sep
mechanical contacts
\sep
atomic scale friction
\sep
boundary lubrication
\sep
relative crystalline orientation
\sep
size scaling
\end{keyword}

\maketitle

\section{Introduction}
\label{introduction}

In a regime of dry friction or boundary lubrication of a single contact,
such as studied by atomic force microscope (AFM) techniques, the friction
force depends substantially and nontrivially on the relative crystalline
orientation of the facing surfaces, as was demonstrated experimentally by
Dienwiebel \textit{et al.}~\cite{DVPFHZ2004}.
Special angles lead to superlubric sliding, but tend to be energetically
unfavorable~\cite{FDFKU2008}.
Depending on the contact mechanical details and the sliding speed, such
superlubric orientations could have long enough time to reconstruct,
approaching a state with a lower free energy but characterized by a higher
barrier (aging), or be retained for long enough for them to provide a
substantial superlubric contribution to the overall sliding dynamics.
The connection between the nanoscale, where friction occurs through the
breaking and formation of local contacts, and the meso- or macroscale,
where many breaking junctions interact elastically, is commonly described
by earthquake-type
models~\cite{cite43,cite44,cite45,FKU2004,BBU2009,BU2010}, or by a master
equation approach~\cite{cite44,BP2008,BP2010,BT2009,SW2009}, or by models
inspired to the Greenwood-Williamson one~\cite{Greenwood66,Greenwood11}
such as the sub-boundary lubrication
model~\cite{Polycarpou98a,Polycarpou98b,Yu04}.
Except the case of ideally flat surfaces such as mica in surface-force
apparatus (SFA) experiments, contact is always realized at microscopic
asperities, whose size ranges typically in the nanometer to micrometer
range.
%
Even when a lubricant is present at the contacting asperities, in a
boundary or sub-boundary lubrication context, the residual lubricant is
often frozen by pressure and confinement, and hydrodynamic viscous sliding
is replaced by a static (stick-slip) contact-breaking regime, which we
focus on in the present work.
In the multiasperity contact, where relative orientation of individual
asperities is not really under control, the most important information to
be extracted from a single-asperity model is the probability distribution
$P_c(\varepsilon_a)$ of the slip activation barriers $\varepsilon_a$.

Molecular dynamics (MD) approaches to lubricated sliding friction~%
\cite{Sorensen96,P0,BN2006,MR2000,HR2001s,HR2001k,Castelli08Lyon,Castelli09,Bonelli09,FasolinoInTrieste,friction_rotate}
are usually forced to use some form of periodic boundary conditions (PBC) in
order to prevent the escape of lubricant atoms (molecules) from the
contacting region under high load, and to keep the simulation size under
control.
PBC mimic the infinite size limit, which might not be especially
appropriate for sub-micrometer--size contacts between sliding surfaces.
Moreover, a fully atomistic model would be computationally too demanding
for a full statistical study of the size
and angular dependency of the characteristics of contacts.

To study the contact, we introduce a simple rigid model for a finite-size
breaking junction realized by the contact of a finite portion of two
different crystalline surfaces.
Such a rigid model could not possibly account for wear or for the
dissipation occurring at contact breaking, as could MD simulations instead,
but provides semi-quantitative estimates of the barriers opposing the onset
of sliding, i.e., the static friction thresholds.
The simple model allows us to evaluate the relevant statistical
distribution of barrier energies.
This distribution enters as a basic ingredient in the master-equation
formulation, which, depending on the general shape of this distribution
leads to different macroscopic sliding regimes, either stick-slip
or smooth-sliding dynamics~\cite{BP2008,BP2010}.

The present paper reports progress beyond our MD study of
Ref.~\cite{friction_rotate}.
For the MD simulations of that work, we used a fixed size of the contact
area, and applied PBC to impose a given commensurability ratio with minimal
boundary effects, and to obtain a fair comparison of different misfit
angles $\phi$.
Of course, the friction force $f$, and in particular its dependency on the
angle $\phi$, does change with the system size.
For a larger contact area, increasingly many incommensurate angles emerge,
producing a more irregular dependency $f(\phi)$.
In the limit of infinite size, $f(\phi)$ develops an infinite number of
singularities~\cite{GT1997}, but this limit is not to be taken too seriously,
since in practice a single contact of a given crystalline orientation never
exceeds a fraction of $\mu$m$^2$ in practical experimental conditions.

A numerical determination of the size dependency of $f(\phi)$ would
constitute a formidable task for MD simulations, since larger sizes are not
only individually more expensive computationally, but would require a
finer and finer sampling of angles $\phi$ and longer equilibration times.
It is instead straightforward to obtain the size dependency of the
rigid-island model.
Thus, \textit{the main goal of the present work is to find}, at least
qualitatively, \textit{the shape of the distribution of static thresholds
  $P_c (\varepsilon_a)$}, which is the \textit{main input of the
  master-equation} approach to friction on meso/macro-scale, for a
\textit{contact of polycrystalline surfaces.}

In Sec.~\ref{model} we spell out the details of this rigid-contact model
for the analysis of the static friction barrier realized by the
contact of a crystalline surface with the boundary lubricant
layer, which we assume in a close-packed ordered configuration.
Sections~\ref{results} and \ref{sizedep} discuss the angle and size
dependency of the friction of a nanocontact, and we compare our results
with those of the MD lubricated model \cite{friction_rotate}.
The basic implications for macroscopic sliding are discussed in
Sec.~\ref{mastereq} within a master-equation approach.
Section~\ref{conclusion} summarizes and discusses our results.

\section{The rigid-island model}
\label{model}

We represent the sliding contact, or the solidified boundary lubricant at
the contact, with a finite rigid crystalline layer of lattice constant
$a$, consisting of $N$ point-like atoms.
We put it in interaction with a substrate potential which is also rigid and
periodic, e.g., a sinusoid of a generally different period $a_s$ and
amplitude $V_0$.
In the case of a one-dimensional system, one can easily find
an explicit expression for the activation energy
barrier for the onset of motion along the chain:
\begin{equation}
\varepsilon_a =
V_0 \left| \frac{ \sin (2\pi N a/a_s) }{ \sin (2\pi a /a_s)} \right|
\,.
\end{equation}
Accordingly, for suitable values of the lattice constant, namely for $a = n
\, a_s /(2 N)$ with integer $n$ (but not a multiple of $N$), this
activation barrier vanishes and the chain moves freely.
For a nonrigid layer, the activation energy remains nonzero for all values
of $a$, but still reaches the first minimum at $a/a_s \propto 1/N$: the
motion in such a case is of a ``caterpillar'' type (for details see
Refs.~\cite{B1990,BraunBook}).

A two-dimensional ``lubricant'' island advancing over the 2D periodic
substrate should exhibit a similar behavior: in particular, a minimum for the
activation energy is expected at $a/a_s \propto 1/ \sqrt{N}$.
The 2D system, however, has one extra degree of freedom, the rotation.
We expect that the activation energy would achieve minima for specific misfit
angles.

To investigate this pattern of minima, we consider a rigid island of size $N$
with a triangular lattice interacting with the simplest 2D substrate
periodic potential
\begin{equation}
V(x,y)= V_0 \, (\sin x + \sin y)
\end{equation}
of square symmetry and lattice spacing $a_s=2\pi$.
The atomic coordinates of the approximately square-shaped island are
$\tilde{x}_{i,j} = X + (i+j/2)a$ and $\tilde{y}_{i,j} = Y + jh$,
where $h=a \sqrt{3}/2$, and the indexes
$i=-j/2,\ldots,-j/2+n_i-1$,
$j=0,\ldots,n_j-1$.
The number of atoms in the rigid flake is $N=n_i n_j$,
with $n_i a \approx n_j h$.
$X$ and $Y$ are the degrees of freedom for the translation of the rigid island.
If we rotate the island by an angle $\phi$,
then the atomic coordinates change to
$x_{i,j} = \tilde{x}_{i,j} \cos \phi - \tilde{y}_{i,j} \sin \phi$ and
$y_{i,j} = \tilde{x}_{i,j} \sin \phi + \tilde{y}_{i,j} \cos \phi$.
For a fixed misfit angle $\phi$, the total potential energy of the
island in contact with the substrate is
\begin{equation}
 U(X,Y)=\sum_{j=0}^{n_j-1}~~ \sum_{i=-j/2}^{n_i-1-j/2} V(x_{i,j},y_{i,j})
\,.
\end{equation}
The set of stationary points of
$U(X,Y)$, defined by
\begin{equation}\label{stationary:eq}
{\partial U}/{\partial X} =0
\;\;\; {\rm and} \;\;\;
{\partial U}/{\partial Y} =0
\,,
\end{equation}
consist of four elements: one minimum $U_m$, two saddle points $U_{s\,1}$
and $U_{s\,2}$, and one maximum.
The nature of the stationary points can be verified by
computing the Hessian of $U(X,Y)$ at each stationary point, but it is in
fact simply provided by the values of the function at the points.
The activation energy for sliding is
\begin{equation}
\varepsilon_a = U_s - U_m
\,,
\end{equation}
where the lower saddle energy $U_s=\min(U_{s\,1},U_{s\,2})$ is considered.
The calculation of the energy barrier $\varepsilon_a$ is extremely fast and
efficient, since it only requires the search of the stationary points in a
2D space, by solving a simple numerical equation (\ref{stationary:eq}).
Given the simplicity of this model, it allows us to compute $\varepsilon_a$
for a very fine sampling of angles and contact sizes, as would be
practically impossible if a fully atomistic simulation is used.
%
Note that the computed barrier height is relevant irrespective of the
direction in which the rigid island is pushed forward.

\begin{figure}
\centerline{\includegraphics[width=8cm,clip=]{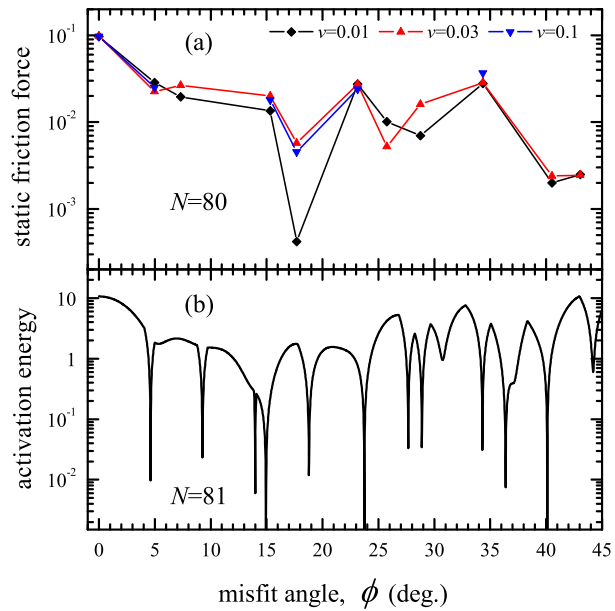}}
\caption{\label{compare}(Color online)
(a) The static friction force (immediately before slip) as a function of
  the misfit angle $\phi$ for a one-layer lubricant film at zero
  temperature for different driving speeds, computed with MD simulations
  with in-plane PBC~\cite{friction_rotate}.  The friction force and driving
  speed are given in natural model units, of the order 1 nN and $\simeq
  1$~m/s respectively.  For full details on these simulations, see
  Ref.~\cite{friction_rotate}.
(b) The activation energy $\varepsilon_a$ (units of $V_0$) for the sliding
  of a rigid island composed of $N=81 =9 \times 9$ lubricant atoms over the
  rigid square substrate as a function of $\phi$.
}
\end{figure}

Although not completely equivalent, the activation energy barrier is indeed
related to the threshold force $f_s$ necessary to initiate sliding.
To compare the results of the present rigid-island approach with the fully
atomistic simulations of Ref.~\cite{friction_rotate}, we take $a /a_s =
4.14/3$, and assume that $f_s = \kappa\, \varepsilon_a /a_s$, where
$\kappa$ is a factor of order unity.
%
This comparison is shown in Fig.~\ref{compare} for an intermediate island
size.
We see that, at least qualitatively, MD and the present simple rigid model
agree on predicting static-friction minima near a set of ``special''
angles.
The singularities at the optimal angles are exact zeroes for the rigid
model: here the two lowest stationary points of $U(X,Y)$ mutate into a
degenerate trough.
%
Unfortunately, within the PBC setup of MD it is impossible to cover such a
fine sampling of misfit angles as allowed by the rigid model one: it is
thus impossible to verify to what extent the two models agree or disagree
on the specific superlubric angular orientations.
The purpose of this comparison is purely to highlight a general qualitative
similarity of the outcomes of two models, which are distinct and address
significantly different physics.
The main difference between the rigid model and the deformable lubricant
film is that the deformable model has a finite barrier against sliding
for all angles, even at optimal ones.

\section{Barrier versus misfit angle}
\label{results}

\begin{figure}
\centerline{\includegraphics[width=8cm,clip=]{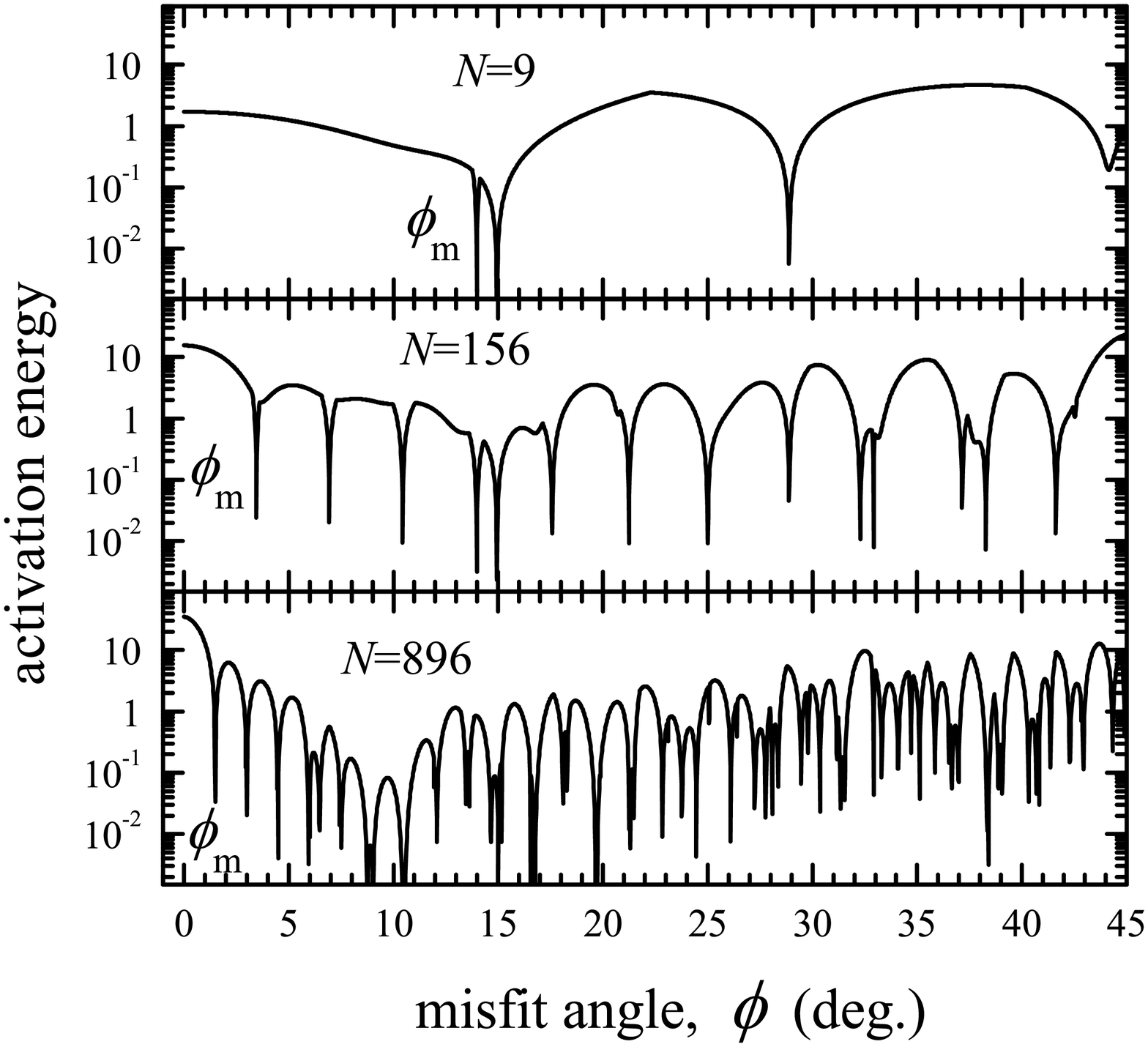}}
\caption{\label{phialpha} 
The activation energy $\varepsilon_a$ (units of $V_0$) as a function of the
misfit angle $\phi$ for three rigid lubricant islands composed by $N=9$,
156 and 896 atoms.
}
\end{figure}

Figure~\ref{phialpha} shows the angular dependency of the barrier against
sliding $\varepsilon_a (\phi)$, for different sizes of the sliding island.
The barrier reaches its first minimum at a misfit angle
$\phi_{\rm m} \approx \pi/(4 \sqrt{N})$, which moves to smaller and smaller
angle as the size $N$ of the island grows.

\begin{figure}
\centerline{\includegraphics[width=8cm,clip=]{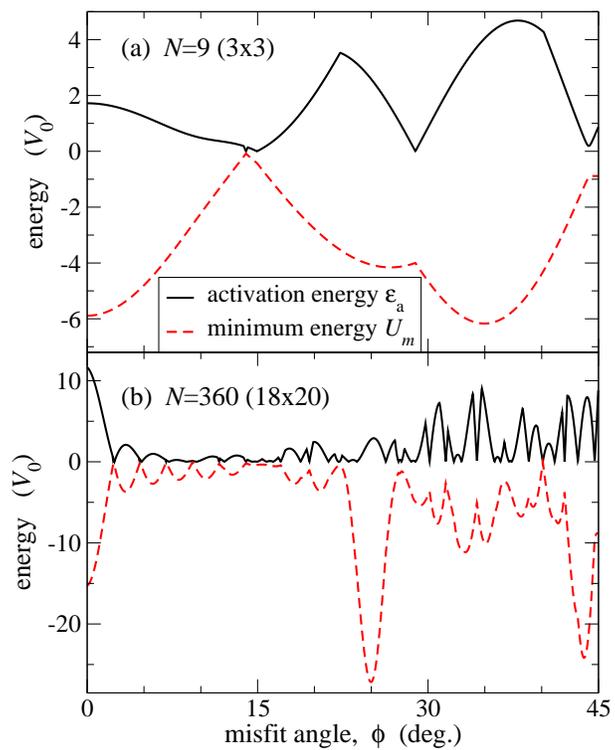}}
\caption{\label{rigid_minen_barrier} (Color online)
A direct comparison of the energy $U_m$ of the minimum (dashed) with the
activation barrier $\varepsilon_a$ (solid), as functions of the misfit
angle $\phi$ for two sizes of the rigid island.
Energies are in units of the substrate potential corrugation $V_0$.
}
\end{figure}

\begin{figure}
\centerline{\includegraphics[width=8cm,clip=]{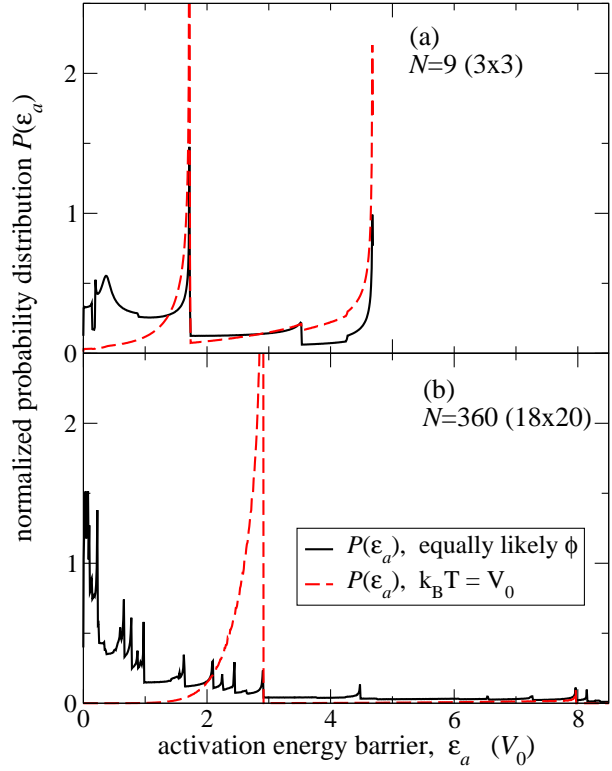}}
\caption{\label{rigid_barrier_distrib}  (Color online)
The distribution of activation barriers $\varepsilon_a$ for two island
sizes reported in the corresponding panels of
Fig.~\ref{rigid_minen_barrier}.
The solid line is computed assuming all misfit angles $\phi$ are equally
likely, the dashed line weights different angles according to a Boltzmann
distribution of the corresponding minimum energy $U_m$, for a
temperature $k_{\rm B} T = V_0$ matching the typical potential barrier of a
single lubricant particle.
}
\end{figure}

Figure~\ref{rigid_minen_barrier} shows that high barrier energies
$\varepsilon_a$ correlate well with rather stable configurations
characterized by a low minimum energy $U_m$, while low-barrier superlubric
angles are usually characterized by unstable (high $U_m$) configurations.
A higher stability of a given angular configuration could make that
configuration more likely, if the asperity is free to rotate.
We take this correlation into account when we evaluate the
distribution of depinning thresholds in the rigid model, based on the
evaluation of $\varepsilon_a$ over a fine grid of angles, as illustrated
by Fig.~\ref{rigid_barrier_distrib}.
For a given contact size $N$, the distribution of activation barriers
exhibits weak divergences produced by the round maxima of
$\varepsilon_a(\phi)$, plus jump discontinuities produced by the ``kinky''
maxima associated to a crossing of the saddle points, as illustrated for
two sizes by the comparison of Figs.~\ref{rigid_minen_barrier} and
\ref{rigid_barrier_distrib}.

If the individual contacts are allowed the freedom and a sufficiently long
time to rotate, thermal fluctuations will lead to geometric relaxation,
eventually leading to an appropriate angular distribution $P_\phi(\phi)$;
if one can neglect the interaction of the contacting grain with the rest of
the slider, this distribution should match a Boltzmann distribution
$P_\phi(\phi)\propto \exp[-U_m (\phi) /k_{\rm B} T]$ of the fully relaxed
energy $U_m$ of the grain-substrate interaction.
If, on the contrary, misfit angles are frozen by the microcrystalline
nature of the surfaces in contact for much longer than the time of the
experiment, all angles are equally likely and $P_\phi(\phi)$ is a constant
(equivalent to the limiting Boltzmann distribution for $T\to \infty$).
Averaging with these two different weight patterns leads to related but
significantly different distributions, as illustrated by the comparison of
dashed and solid lines in Fig.~\ref{rigid_barrier_distrib}.
Observe in particular that the effect of the Boltzmann weights is to
suppress the probability of small activation barriers $\varepsilon_a$: this
is a consequence of the stable angles (minima of $U_m$) being typically
associated to high barriers $\varepsilon_a$, as remarked above (see
Fig.~\ref{rigid_minen_barrier}).
If the atomic layer represents a frozen lubricant, then one should beware
of other $\phi$-dependent energy contributions to be added to $U_m$ due to
the interaction with the crystalline anisotropy of the asperity region of
the upper slider.
These extra terms would of course influence the Boltzmann weights in the
fast-rotating condition, in a way which could only be predicted in a
condition where the details of this interaction and relative crystalline
alignment were given.

\section{The distribution of static thresholds}
\label{sizedep}

\begin{figure}
\centerline{\includegraphics[width=8cm,clip=]{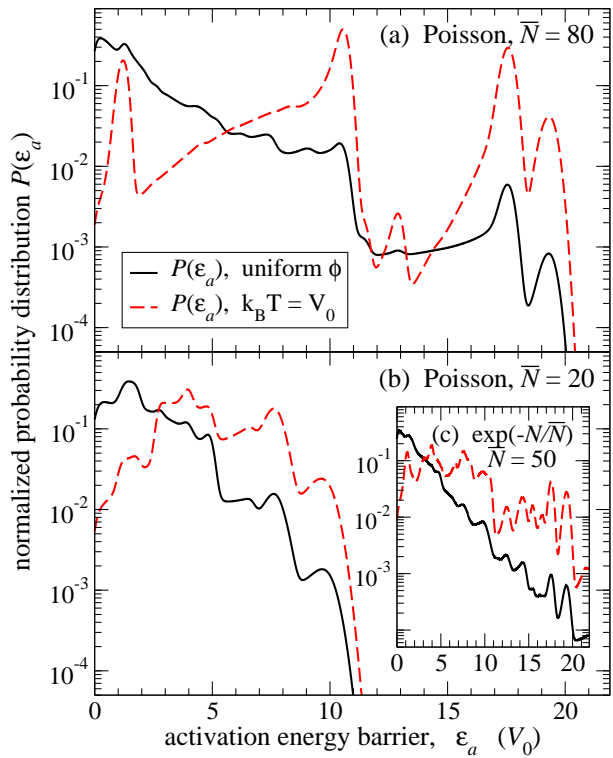}}
\caption{\label{rigid_barrier_total_distrib} (Color online)
The probability distribution of static thresholds for the rigid domains
averaged over the domain size $N$ with the Poissonian distribution of
average domain size (a) $\bar{N} =80$ and (b) $\bar{N} =20$.
Inset (c): the probability distribution computed by averaging over an
exponential size distribution $\exp(-N/\bar{N})$ with $\bar{N}=50$.
Averaging over the misfit angle $\phi$ is carried out
for solid (all angles are presented equally likely) and
dashed (different angles are weighted according to the Boltzmann distribution)
lines as in Fig.~\ref{rigid_barrier_distrib}.
}
\end{figure}

To describe friction in a meso- or macroscopic multi-contact regime it makes sense
to assume a distribution of contact sizes $N$, and obtain the statistical
contact properties by averaging over $N$.
For the size distribution one may take a Poisson distribution
$P(N) = (\bar{N}^N/N!) \, \exp(-\bar{N}) $
with an average domain size $\bar{N}$.
We include clusters of nearly squared shape only.

By combining this size distribution with the distributions of unpinning
barriers for individual sizes $N$, calculated in the previous Section, we
obtain a global distribution of barrier heights.
The resulting distribution is displayed in
Fig.~\ref{rigid_barrier_total_distrib}, where we have smoothened the
singularities of the distribution for each individual size $N$ by means of
the convolution with a Gaussian of full width at half maximum matching the
average inter-peak spacing (varying from $0.1\,V_0$ to $2.5\,V_0$).
We see that this distribution decays roughly exponentially by approximately
two decades, and then drops rapidly due to the fast large-size decay of
the Poisson function combined with the decreasing probability of
barriers of increasing height, as illustrated in
Fig.~\ref{rigid_barrier_distrib}b.
The choice of a Poisson distribution is not especially critical: a similar
distribution of static thresholds is obtained if we assume an exponential
size distribution $P(N) = \bar{N}^{-1} \exp (-N/\bar{N})$, see
Fig.~\ref{rigid_barrier_total_distrib}c.

\begin{figure}
\centerline{\includegraphics[width=8cm,clip=]{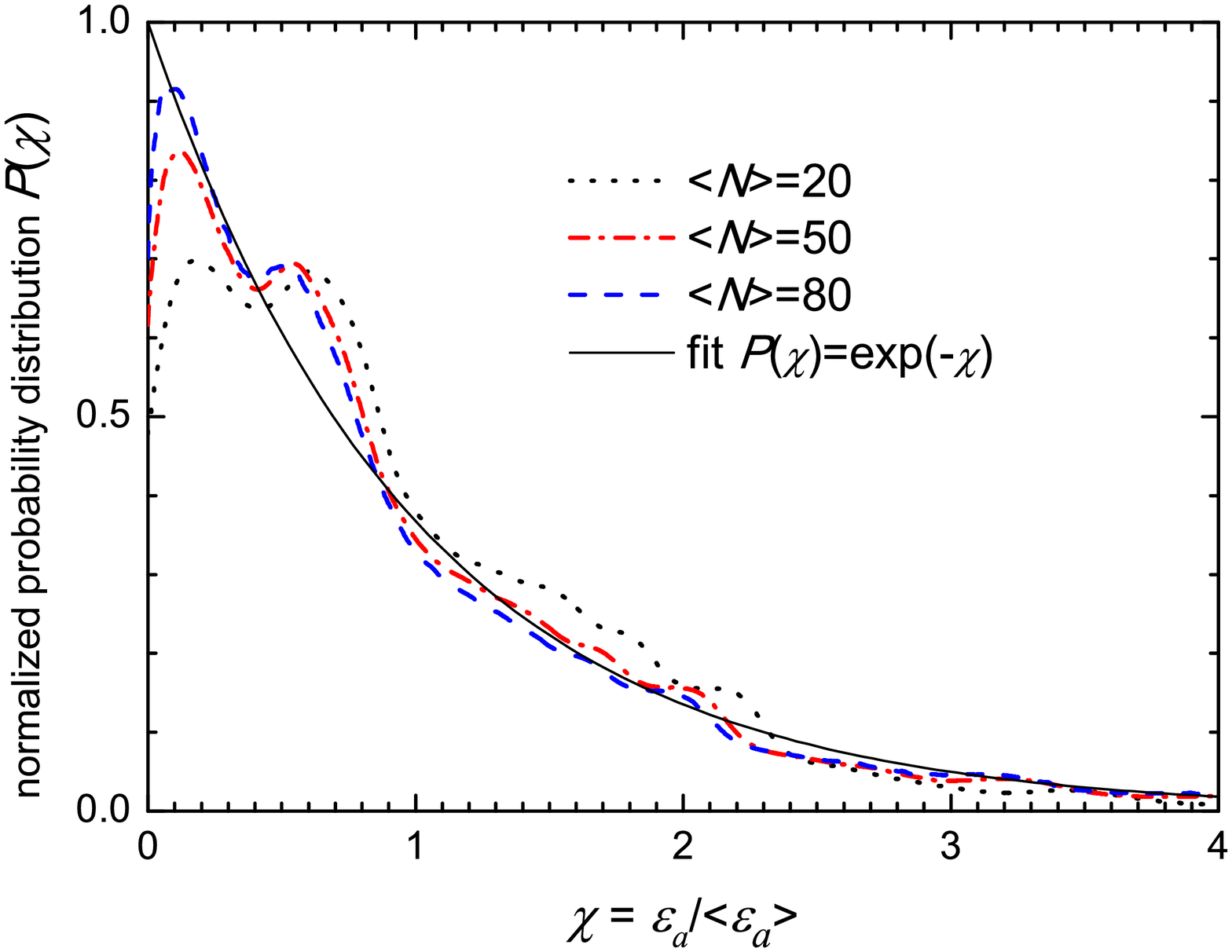}}
\caption{\label{C18a} (Color online)
The probability distribution of static thresholds $P_c (\chi)$ as a
function of renormalized barrier heights $\chi = \varepsilon_a / \langle
\varepsilon_a \rangle$ for the rigid domains averaged over the domain size
$N$ with an exponential size distribution $\exp(-N/\bar{N})$ with
$\bar{N}=20$ (dotted/black), 50 (dot-dashed/red) and 80 (dashed/blue).
}
\end{figure}

The comparison of panels (a) and (b) of
Fig.~\ref{rigid_barrier_total_distrib} shows that the average contact size
$\bar N$ affects the quantitative detail of the distribution, but not its
qualitative shape.
Moreover, if we plot the individual distributions as functions of the
dimensionless rescaled activation energy barrier
\begin{equation}
\chi = \varepsilon_a / \langle \varepsilon_a \rangle
\,,
\end{equation}
all distributions may be roughly approximated by the exponential function
$P_c (\chi) = \exp (- \chi )$ as demonstrated in Fig.~\ref{C18a}.
Remarkably, the general shape of the distribution of barriers of the
rigid-island model resembles the distribution, shown in Fig.~9 of
Ref.~\cite{friction_rotate}, of static thresholds in the lubricated model
based on PBC and a single size $N=80$.
The main visible difference is that, for the deformable domains of
Ref.~\cite{friction_rotate}, the probability of small barriers is
significantly lower than for the rigid slider at hand.

\begin{figure}
\centerline{\includegraphics[width=8cm,clip=]{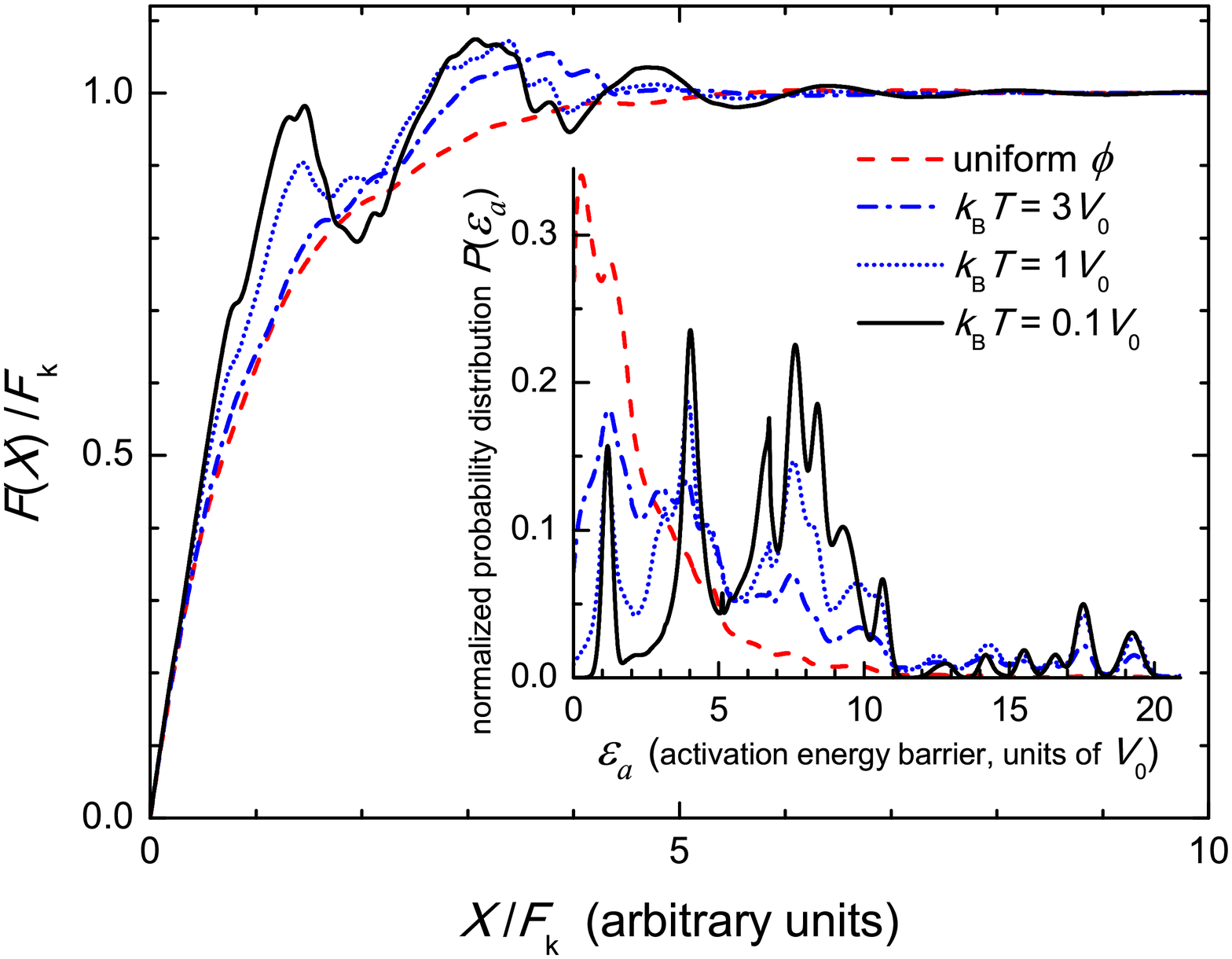}}
\caption{\label{C18b} (Color online)
The friction force $F(X)$ (normalized on the $X \to \infty$ limiting value
$F_k$, corresponding to the kinetic friction force in the smooth-sliding
regime) as a function of the displacement $X$ of the rigid slider for
different distributions of static thresholds shown in inset.
Inset: the probability distribution of static thresholds for the rigid
domains averaged over the domain size $N$ with an exponential size
distribution $\exp(-N/\bar{N})$ with $\bar{N}=50$, when averaging over the
misfit angle $\phi$ is weighted according to a Boltzmann distribution of
the domain energy, for different temperatures.
}
\end{figure}

Small barriers $\varepsilon_a$ become also suppressed when the islands
can rotate so that angles are distributed thermally.
As temperature decreases, the distribution of static thresholds exhibits
more and more pronounced local maxima at values corresponding to minima of
the domain's potential energy (see inset of Fig.~\ref{C18b}).

\section{Consequence for macroscopic sliding}
\label{mastereq}

Once the distribution of static thresholds is known, we can predict the
dynamics of the tribological system with the help of
the master-equation approach based on an
earthquake-like model~\cite{BP2008,BP2010,BT2009}.
Consider the contact of two rough unlubricated substrates (top and bottom)
on a meso- or macroscale.
If the position of the bottom substrate is kept fixed and the top substrate
is displaced by a distance $X$, the interface force (shear stress) begins
to grow, $F \propto X$.
Then, the domains where the stress exceeds a corresponding
threshold value, start to slide, thus relaxing the local stress,
and the increase of the total force $F$ will degrade.
The overall average dependence $F(X)$ follows from a solution of this
master equation, where the distribution of static thresholds is the input
parameter, as discussed in detail in Refs.~\cite{BP2008,BP2010,BT2009}.

For the threshold distributions calculated above, typical dependences
$F(X)$ are shown in Fig.~\ref{C18b}.
When the misfit angles are distributed equally likely so that the
distribution of static thresholds is nonzero down to zero threshold and
$P_c(\varepsilon_a)$ has no sharp maxima, the force $F$ increases
monotonically with $X$, approaching the kinetic-friction force $F_k$
characteristic of smooth sliding for $X \to \infty$.
Thus, in such kind of macroscopic slider, static and kinetic friction
forces coincide, and the motion always corresponds to smooth sliding.
In contrast, when the threshold distribution exhibits well pronounced sharp
maxima, like for thermalized domains or for an ordered homogeneous thin
lubricant film, the function $F(X)$ reaches a maximum (which represents the
macroscopic static friction force $F_s$) greater than $F_k$, and then
decreases as the asperities give way collectively.
For a soft enough slider, the inequality $dF(X)/dX < 0$ for some $X$ leads
to the appearance of the elastic instability
in the system dynamics~\cite{BP2008,BP2010}.
Under such conditions, the macroscopic slider may exhibit stick-slip
motion, provided the slider is soft and the delay in contact reformation
is taken into account~\cite{BT2009}.

\section{Discussion and Conclusion}
\label{conclusion}

The calculations within the simple model at hand provide, first of all,
relevant insight for a single-asperity microscopic system, e.g.\ for a
flake sliding over an atomically flat surface, where the slider may rotate
and search for a local minimum of the potential energy.
As a consequence, even if sliding starts off in a low-friction state (e.g.,
in the superlubric state associated to a highly incommensurate $\phi$
\cite{DVPFHZ2004}), the flake will eventually rotate and spend most of its
time near a local minimum (Fig.~\ref{rigid_minen_barrier} indicates that
for the triangle-on-square geometry of the present model such minimum need
not be $\phi =0$).
Accordingly, after a relaxation time typical of the flake rotation,
friction should increase (as predicted by the low-$\varepsilon_a$ side drop
of the distribution of Fig.~\ref{rigid_barrier_total_distrib}).
Such a behavior was observed experimentally and in MD
simulation~\cite{FDFKU2008}.
Observe also that an increased rate of thermally activated jumps across the
pinning barriers would additionally lead to a thermolubric
regime~\cite{Jinesh08,Krylov08}.

More than single-asperity experiments, the focus of the present work
concerns meso- and macroscopic sliding friction.
At the nanoscopic level, the friction force produced by a sliding contact
depends substantially and nontrivially on the relative crystalline
orientation of the facing surfaces.
In the present work we provide a basic tool to connect between the
nanoscale, where friction occurs through the breaking and formation of
local contacts, and the meso/macro-scale, where many breaking junctions
interact elastically, as commonly described by an earthquake-type model or
by a master-equation approach.
The quantity that summarizes the information obtained by averaging over all
possible contact sizes and angles is a probability distribution
$P_c(\varepsilon_a)$ of the slip activation barriers $\varepsilon_a$.
Our simple model permits us to evaluate such a distribution of barrier
energies, reaching beyond the small sizes and few rotation angles allowed
by detailed microscopical MD simulations.
This distribution is a basic ingredient for the master-equation
formulation, which, depending on the actual shape of this distribution can
lead to different general macroscopic sliding regimes.
The analysis of the shape of this distribution allows one to understand the
physics of the meso-macroscopic sliding in terms of the underlying
microscopic junction-breaking statistical properties.

Two basic regimes of macroscopic sliding emerge from this model:
(\textit{i}) When superlubric alignments are suppressed by aging to
thermodynamically more favorable alignments, a nonmonotonic peaked
distribution $P_c(\varepsilon_a)$ of barrier heights is obtained, which
tends to induce a macroscopic stick-slip regime.
(\textit{ii}) In contrast, when the probability of weak activation barriers
is sufficiently large to produce a monotonically decaying distribution
$P_c(\varepsilon_a)$, then macroscopic smooth sliding is possible, even in
the presence of microscopic breaking-junction dynamics.

The present simple and very idealized model is not meant to address any
specific properties of a well-defined contacting system, but it focuses on
the possibility to extract macroscopic statistical information out of the
mechanical properties of contacts.
Many details of real contacts are left out, including surface curvature,
wear, local thermal expansion.
For this reason, it would be interesting (although extremely expensive
numerically) to attempt a similar statistical method using the MD
simulations of a specific contact described in terms of realistic force
fields and curved surfaces.
While the quantitative detail of $P_c(\varepsilon_a)$ is likely to depend
on the specific contacting materials, its general properties should mostly
follow those determined by means of the present simple model.


\section*{Acknowledgments}

We wish to express our gratitude to B.N.J.\ Persson for helpful
discussions.
This research was supported in part by a grant from the Cariplo Foundation
managed by the Landau Network -- Centro Volta,
and by the
Italian National Research Council (CNR, contract ESF/\-EUROCORES/\-FANAS/\-AFRI),
whose contributions are gratefully acknowledged.



\begin{thebibliography}{99}

\bibitem{DVPFHZ2004}
  M.\ Dienwiebel, G.S.\ Verhoeven, N.\ Pradeep, J.W.M.\ Frenken,
  J.A.\ Heimberg, and H.W.\ Zandbergen,
  Phys.\ Rev.\ Lett.\  {\bf 92}, 126101 (2004).
\bibitem{FDFKU2008}
  A.E.\ Filippov, M.\ Dienwiebel, J.W.M.\ Frenken,
  J.\ Klafter, and M.\ Urbakh,
  Phys.\ Rev.\ Lett.\  {\bf 100}, 046102 (2008).
\bibitem{cite43}
  Z.\ Olami, H.J.S.\ Feder, and K.\ Christensen,
  Phys.\ Rev.\ Lett.\ {\bf 68}, 1244 (1992).
\bibitem{cite44}
  B.N.J.\ Persson,
  Phys.\ Rev.\ B {\bf 51}, 13568 (1995).
\bibitem{cite45}
  O.M.\ Braun and J.\ R\"oder,
  Phys.\ Rev.\ Lett.\ {\bf 88}, 096102 (2002).
\bibitem{FKU2004} 
  A.E.\ Filippov, J.\ Klafter, and M.\ Urbakh,
  Phys.\ Rev.\ Lett.\  {\bf 92}, 135503 (2004).
\bibitem{BBU2009} 
  O.M.\ Braun, I.\ Barel, and M.\ Urbakh,
  Phys.\ Rev.\ Lett.\  {\bf 103}, 194301 (2009).
\bibitem{BU2010} 
  I.\ Barel, M.\ Urbakh, L.\ Jansen, and A.\ Schirmeisen,
  Phys.\ Rev.\ Lett.\  {\bf 104}, 066104 (2010).
\bibitem{BP2008}
  O.M.\ Braun and M.\ Peyrard,
  Phys.\ Rev.\ Lett.\  {\bf 100}, 125501 (2008).

\bibitem{BP2010}
  O.M.\ Braun and M.\ Peyrard, Phys.\ Rev.\ E {\bf 82}, 036117 (2010).

\bibitem{BT2009}
  O.M.\ Braun and E.\ Tosatti,
  Europhys.\ Lett.\ {\bf 88}, 48003 (2009).
\bibitem{SW2009}
  M.\ Srinivasan and S.\ Walcott,
  Phys.\ Rev.\ E  {\bf 80}, 046124 (2009).

\bibitem{Greenwood66} 
J.A. Greenwood, J.B.P. Williamson, Proc. R. Soc. Lond. A {\bf 295}, 300 (1966).

\bibitem{Greenwood11} 
J.A. Greenwood, C. Putignano, and M. Ciavarella, Wear {\bf 270}, 332 (2011).

\bibitem{Polycarpou98a} 
A. A. Polycarpou and Izhak Etsion, J. Tribol. {\bf 120}, 296 (1998).

\bibitem{Polycarpou98b} 
A. A. Polycarpou and I. Etsion, Tribol. Trans. {\bf 41}, 217 (1998).

\bibitem{Yu04} 
N. Yu, S. R. Pergande, and A. A. Polycarpou,
J. Tribol. {\bf 126}, 626 (2004).

\bibitem{Sorensen96}
  M. R. S\o{}rensen, K. W. Jacobsen, and P. Stoltze,
  Phys. Rev. B \textbf{53}, 2101 (1996).
\bibitem{P0}
  B.N.J.\ Persson,
  {\em ``Sliding Friction: Physical Principles and Applications''}
  (Springer-Verlag, Berlin, 1998).
\bibitem{BN2006}
  O.M.\ Braun and A.G.\ Naumovets,
  Surf.\ Sci.\ Reports {\bf 60}, 79 (2006).
\bibitem{MR2000}
  M.H.\ M\"user and M.O.\ Robbins,
  Phys.\ Rev.\ B  {\bf 61}, 2335 (2000).
\bibitem{HR2001s}
  G.\ He and M.O.\ Robbins,
  Phys.\ Rev.\ B  {\bf 64}, 035413 (2001).
\bibitem{HR2001k}
  G.\ He and M.O.\ Robbins,
  Tribology Lett.\ {\bf 10}, 7 (2001).
\bibitem{Castelli08Lyon}
  I. E. Castelli, N. Manini, R. Capozza, A. Vanossi,
  G. E. Santoro, and E. Tosatti,
  J. Phys.: Condens. Matter {\bf 20}, 354005 (2008).
\bibitem{Castelli09} 
  I. E. Castelli, R. Capozza, A. Vanossi, G.E. Santoro,
  N. Manini, and E. Tosatti,
  J. Chem. Phys. {\bf 131}, 174711 (2009).
\bibitem{Bonelli09}
  F.\ Bonelli, N.\ Manini, E.\ Cadelano, and L.\ Colombo,
  Eur.\ Phys.\ J.\ B {\bf 70}, 449 (2009).
\bibitem{FasolinoInTrieste}
  A.S.\ de Wijn, C.\ Fusco, and A.\ Fasolino,
  Phys.\ Rev.\ E  {\bf 81}, 046105 (2010).
\bibitem{friction_rotate}
  O.M.\ Braun and  N.\ Manini, Phys.\ Rev.\ E {\bf 83}, 021601 (2011). 
%
\bibitem{GT1997}
  T.\ Gyalog and H.\ Thomas,
  Europhys.\ Lett.\ {\bf 37}, 195 (1997).
%
\bibitem{B1990}
  O.M.\ Braun,
  Surface Sci.\ {\bf 230}, 262 (1990).
\bibitem{BraunBook}
  O.M.\ Braun and Yu.S.\ Kivshar,
  {\em ``The Frenkel-Kontorova Model: Concepts, Methods, and Applications''}
  (Springer-Verlag, Berlin, 2004).
%
\bibitem{Jinesh08}
  K.B.\ Jinesh, S.Yu.\ Krylov, H.\ Valk, M.\ Dienwiebel, and J.W.M.\ Frenken,
  Phys.\ Rev.\ B  {\bf 78}, 155440 (2008).
\bibitem{Krylov08}
  S. Yu. Krylov and J. W. M. Frenken,
  J.\ Phys.: Condens.\ Matter {\bf 20}, 354003 (2008).

\end{thebibliography}
\end{document}